\documentclass[apsrev,prl,twocolumn,superscriptaddress]{revtex4-1}
\usepackage{amssymb,amsfonts,amsmath}
\usepackage{adjustbox}
\usepackage{hyperref}
\usepackage{color,ulem}
\makeatletter

\begin{document}
\title{Group-IV monochalcogenide monolayers: two-dimensional ferroelectrics with weak intra-layer bonds and a phosphorene-like monolayer dissociation energy}
\author{Shiva P. Poudel}
\affiliation{Department of Physics, University of Arkansas, Fayetteville, Arkansas 72701, United States}
\author{John W. Villanova}
\affiliation{Department of Physics, University of Arkansas, Fayetteville, Arkansas 72701, United States}
\author{Salvador Barraza-Lopez}
\email{sbarraza@uark.edu}
\affiliation{Department of Physics, University of Arkansas, Fayetteville, Arkansas 72701, United States}
\affiliation{Institute for Nanoscience and Engineering, University of Arkansas, Fayetteville, Arkansas 72701, USA}


\begin{abstract}
We performed density functional theory calculations with self-consistent van der Waals corrected exchange-correlation (XC) functionals to capture the structure of black phosphorus and twelve monochalcogenide monolayers and find the following results: (a) The in-plane unit cell changes its area in going from the bulk to a monolayer. The change of in-plane distances implies that bonds weaker than covalent or ionic ones are at work within the monolayers themselves. This observation is relevant for the prediction of the critical temperature $T_c$. (b) There is a hierarchy of independent parameters that uniquely define a ground state ferroelectric unit cell (and square and rectangular paraelectric unit cells as well): only 5 optimizable parameters are needed to establish the unit cell vectors and the four basis vectors of the ferroelectric ground state unit cell, while square and rectangular paraelectric structures are defined by only 3 or 2 independent optimizable variables, respectively. (c) The reduced number of independent structural variables correlates with larger elastic energy barriers on a rectangular paraelectric unit cell when compared to the elastic energy barrier of a square paraelectric structure. This implies that $T_c$ obtained on a structure that keeps the lattice parameters fixed (for example, using an NVT ensemble) should be larger than the transition temperature on a structure that is allowed to change in-plane lattice vectors (for example, using the NPT ensemble). (d) The dissociation energy (bulk cleavage energy) of these materials is similar to the energy required to exfoliate graphite and MoS$_2$. (e) There exists a linear relation among the square paraelectric unit cell lattice parameter and the lattice parameters of the rectangular ferroelectric ground state unit cell.  These results highlight the subtle atomistic structure of these novel 2D ferroelectrics.
\end{abstract}

\maketitle

\section{Introduction}

Group-IV monochalcogenide monolayers \cite{tritsaris_jap_2013_sns,singh_apl_2014_ges_gese_sns_snse} are ferroelectric \cite{fei_apl_2015_ges_gese_sns_snse} semiconducting membranes two atoms thick that exist experimentally already \cite{Kai,KaiPRL}. The lack of inversion symmetry encoded in their in-plane ferroelectric moment underpins their nonlinear optical properties \cite{b2,b4,Wangeaav9743}. Presently, digital storage relies on ferromagnetic memories but these ultrathin ferroelectrics hold promise for all-electric non-volatile memories, too \cite{kai3}. These materials of extremely small thickness undergo a phase transformation from an average rectangular ferroelectric configuration onto an average square paraelectric configuration \cite{Mehboudi2016,other2,Kai} at finite temperature \cite{Mehboudi2016}.

Despite their potential, there exists an unusually wide range of lattice parameters $a_{1,0}$ and $a_{2,0}$  reported for these monolayers \textit{even when using exactly identical computational tools} \cite{tomanek_acsnano_2015_sis,tuttle_prb_2015_sis,jiang_scibull_2017_sis_sise,mao_cpl_2018_sise,chen_jmcc_2016_site,wang_nanolett_2017_gese,wu_prb_2017_all,fei_prl_2016,
wang_2dmat_2017_mmls,kamal_prb_2016_iv_vi_monolayers,chowdhury_jpc_2017_mmls,qin_nanoscale_2016_ges_gese_sns_snse,gomes_prb_2015a_ges_gese_sns_snse,wu_nanolett_2016_ges_gese_sns_snse,
huang_jcp_2016_ges_gese_sns_snse,fei_apl_2015_ges_gese_sns_snse,singh_apl_2014_ges_gese_sns_snse,shafique_scireport_2017_ges,liu_prl_2018_ges,
panday_jpcm_2017_ges_gese_sns_snse,rangel_prl_2017,gomes_prb_2016_excitons,slawinska_2d_mat_2018_mmls,cook_natcom_2017_ges,gomes_prb_2015b_ges_gese_sns_snse,
guo_apl_2017_sns_snse_ges_gese,zhang_nanotech_2016_ges,Mehboudi2016,shen_2dmater_2018_gese,hanakata_prb_2016_sns_gese,hu_apl_2015_gese,other2,lebedev_jap_2018_sns,
haleoot_prl_2017_sns_snse,rodin_prb_2016_sns,seifert_jpcc_2016_p_sns,tritsaris_jap_2013_sns,wang_nanoscale_2015_snse,shi_nl_2015_snse_gese,other4,
hu_nanoscale_2017_snse,Kai,liu_prl_2018_snte,sntebl} (as documented in the Supplementary Material). Such lack of agreement implies that determining the atomistic structure of these materials accurately may not be a trivial endeavor, and it prompts a number of structure-related questions: (i) what is the smallest number of independent variables needed to specify such structures? (ii) how many structural degeneracies do these materials have? (iii) how do interatomic distances change in turning from the bulk into a monolayer phase? (iv) what is their cleavage energy? (v) speaking of monolayers, what is the relation among the lattice parameter of a paraelectric square unit cell $a_s$ and the lattice parameters of the ground state ferroelectric one $a_{1,0}$ and $a_{2,0}$? (vi) what is the elastic energy barrier between ferroelectric and paraelectric monolayers? And finally, is the usual rationale to include dispersive corrections of the electronic structure for bulk layered materials, yet eschew them for isolated monolayers, always a sound assumption, even in these soft monochalcogenide monolayers?

To answer these questions, geometries were determined from density-functional theory \cite{martin} using eight different exchange-correlation (XC) functionals that include traditional ones (LDA \cite{LDA,LDA2} and PBE \cite{PBE}), five with self-consistent van der Waals corrections \cite{Becke,dion2004,reviewvdw} (optPBE-vdW \cite{klimes1,klimes2}, optB86b-vdW \cite{klimes1,klimes2}, vdW-DF-cx \cite{cx}, vdW-DF2 \cite{DF2}, B86R-vdW-DF2 \cite{DF2,Hamada}) and the recently developed SCAN+rVV10 \cite{scan}, which has been successful to describe the weak bonding in liquid and solid water in the most precise manner yet \cite{PNASscan}. To appreciate how dramatically sensitive the structure of group-IV monochalcogenides is to its environment, graphite and a typical transition metal dichalcogenide (MoS$_2$) were studied with these eight functionals for comparison purposes.

Calculations were performed with the VASP code \cite{vasp} (release 5.4.4) on a 30$\times$30$\times$1 $k-$point mesh and with a 600 eV energy cutoff. Energy and force convergence criteria were set to $10^{-11}$ eV and $10^{-5}$ eV/\AA{} respectively, and the high precision tag was turned on. For freestanding monolayer calculations, the out-of-plane lattice vector length was 30 \AA{}. The anharmonicity of the energy landscape of monolayers makes it difficult for standard algorithms that optimize lattice vectors to find the overall minima. We have therefore performed calculations on preestablished lattice parameter meshes in which the variation of energy against lattice parameters is sampled with a 0.005~\AA{} resolution and the four basis atoms are allowed to move along the $x-$ and $z-$directions.

The manuscript is organized as follows: We discuss the structure of the ferroelectric ground state and two paraelectric structures first, expressing their lattice and basis vectors as well as interatomic distances in terms of the smallest set of independent variables. Then, we compare these interatomic distances in the bulk and in the monolayer, to find out that second and third nearest neighbors increase their separation by up to 20\% depending on the mean atomic number (which will be defined momentarily). Such extreme variation of the bond length implies that these neighbors do not form covalent nor ionic bonds but weaker ones. A monolayer dissociation energy is defined next and shown not to be that different from its magnitude in graphite,
MoS$_2$, or black phosphorus, which means that these monolayers should exfoliate readily. The work ends by introducing a relation among the lattice constant $a_s$ of the square
paraelectric unit cell and the lattice constants of the ferroelectric ground state $a_{1,0}$ and $a_{2,0}$, and by comparing the elastic energy barriers to reach paraelectric
square or rectangular structures.

\section{Results and discussion}

\subsection{Ferroelectric ground state and two paraelectric structures}

\begin{figure}[t]
\begin{center}
\includegraphics[width=0.48\textwidth]{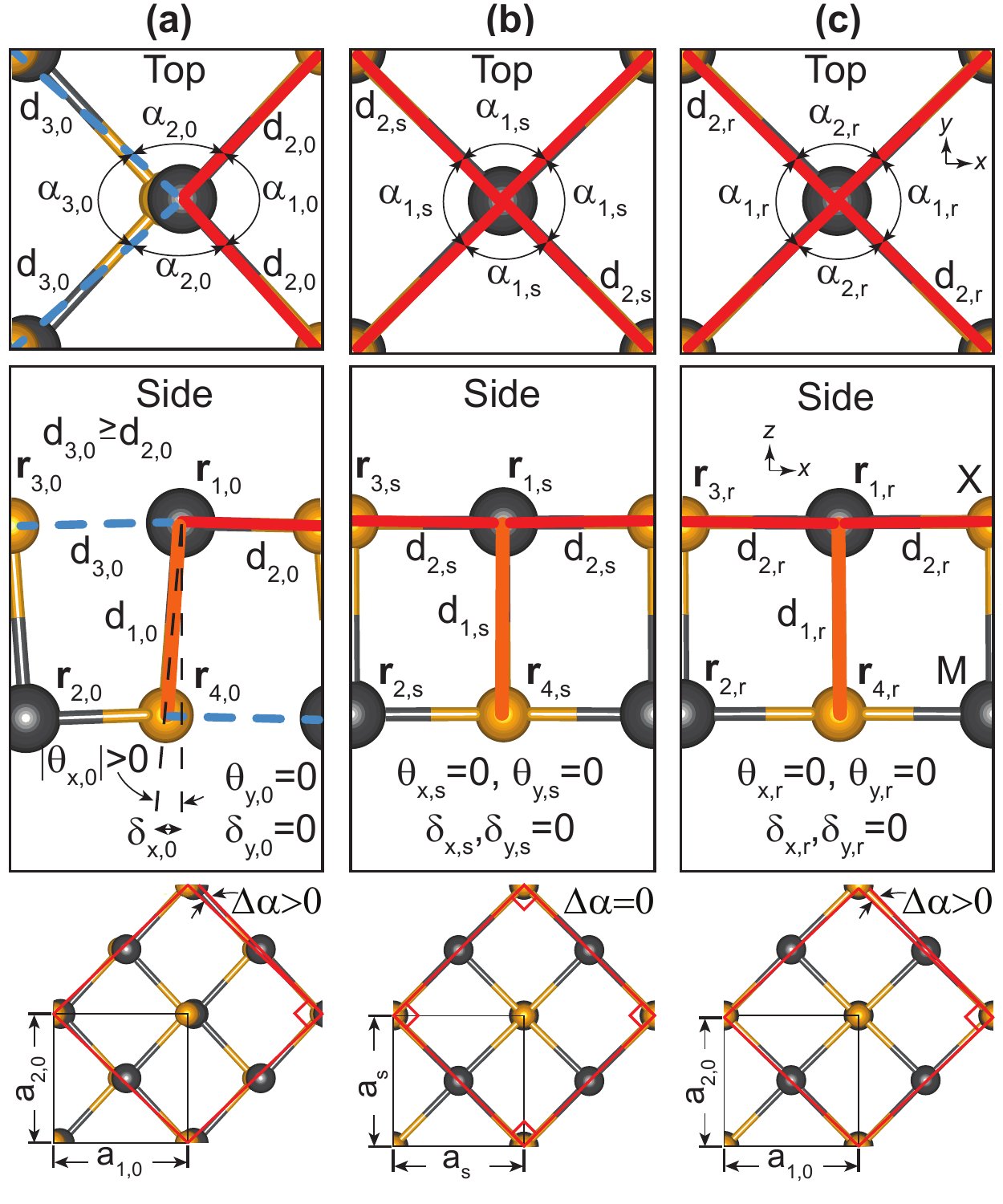}
\caption{(a) Ground state geometry of a group-IV monochalcogenide monolayer. The M atom (Si, Ge, Sn, or Pb) is gray and the X atom (S, Se, or Te) orange (the rhombic distortion angle $\Delta \alpha$ is nonzero). (b) Paraelectric ($\delta_{x,s}=\delta_{y,s}=0$) square phase in which $a_1=a_2=a_s$, $d_{3,s}=d_{2,s}$, and $\alpha_2=\alpha_3=\alpha_{1,s}$ ($\Delta \alpha=0$). (c) Paraelectric ($\delta_{x,r}=\delta_{y,r}=0$) rectangular phase in which $a_1$ and $a_2$ retain their ground-state magnitudes, $d_{3,r}=d_{2,r}$, and $\alpha_3=\alpha_{1,r}$ but $\alpha_{2,r}\ne \alpha_{1,r}$ ($\Delta \alpha > 0$). See Eqs.~(\ref{eq:eq1}-\ref{eq:eq3}).}\label{fig:fig1}
\end{center}
\end{figure}

Their zero-temperature geometry is key to understand these materials' finite temperature behavior. Figures~\ref{fig:fig1}(a) and \ref{fig:fig1}(b) show a SnSe monolayer (representing a typical group-IV monochalcogenide monolayer) in one of its four ground state configurations at zero temperature. Its rectangular unit cell has four basis atoms ($\mathbf{r}_i$, i=1,2,3,4). Gray (dark) atoms ($M$) are a group IV element, and orange (light) atoms ($X$) are a chalcogen heavier than oxygen. (A non-polar, isostructural black phosphorus monolayer was included in this work for comparison purposes, and one should replace the term ferroelectric to \textit{ferroelastic}, and paraelectric to \textit{paraelastic} when this non-polar material is alluded to in what follows.)

Atomic coordinates of the ground state unit cell depend on \textit{five} independent variables: lattice parameters $a_1$ and $a_2$, a horizontal tilt $\delta_x$, and two atomic heights $z_1$ and $z_3$ relative to atom $\mathbf{r}_2$ (which is set at a height $z_2=0$) for a membrane thickness of less than 3 \AA{}. The relative height of atom $\mathbf{r}_4$ (written as an independent parameter in Ref.~\cite{other4}) is $z_4=z_1-z_3$. This way, the structural energy landscape \cite{wales2003energy} belongs to a five-dimensional space $E=E(a_1,a_2,\delta_x,z_1,z_3)$ and the ground state energy is $E_0=min\{E(a_1,a_2,\delta_x,z_1,z_3)\}=E(a_{1,0},a_{2,0},\delta_{x,0},z_{1,0},z_{3,0})$. Lattice and basis vectors are:
\begin{align}
\label{eq:eq1}
{\bf a}_{1,0}=(a_{1,0},0,0),\quad {\bf a}_{2,0}=(0,a_{2,0},0),\qquad\\
\begin{split}
{\mathbf r}_{1,0}=(\tfrac{a_{1,0}}{2}+\delta_{x,0},\tfrac{a_{2,0}}{2},z_{1,0}),\quad {\bf r}_{2,0}=(\delta_{x,0},0,0) &\quad\text{(M)},\\
{\bf r}_{3,0}=(0,0,z_{3,0}),\quad {\bf r}_{4,0}=(\tfrac{a_{1,0}}{2},\tfrac{a_{2,0}}{2},z_{1,0}-z_{3,0}) &\quad\text{(X)}. \nonumber
\end{split}
\end{align}
Structural degeneracies are determined as follows: an effective dipole moment \cite{fei_apl_2015_ges_gese_sns_snse} along the horizontal ($x$) direction is created when $\delta_{x,0}$ is positive. An isoenergetic structure results by reversing the direction of the dipole moment, $\delta_{x,0} \to -\delta_{x,0}$. Swapping $x-$ and $y-$coordinates renders two additional degenerate ground state structures (in isoenergetic horizontal and vertical configurations similar to those discussed for model 2D tissue \cite{marchetti1}) for a total of four degeneracies \cite{Mehboudi2016}. A mirror reflection across the $xy-$plane (or an equivalent displacement by $\mathbf{a}_{1,0}/2$) is what renders $z_{4,0}=z_{1,0}-z_{3,0}$; this additional degeneracy does not produce a new dipole orientation and is usually neglected for that reason.

Thermally-induced changes on geometrical order parameters such as interatomic distances and angles can flag 2D transformations at finite temperature. In the ground state unit cell, first nearest neighbor atoms are separated by a distance $d_{1,0}=|\mathbf{r}_{1,0}-\mathbf{r}_{4,0}|=|\mathbf{r}_{2,0}-\mathbf{r}_{3,0}|=\sqrt{\delta_{x,0}^2+z_{3,0}^2}$, second nearest neighbors are at a distance  $d_{2,0}=|\mathbf{r}_{4,0}-\mathbf{r}_{2,0}|
=\sqrt{(\frac{a_{1,0}}{2}-\delta_{x,0})^2+(\frac{a_{2,0}}{2})^2+(z_{1,0}-z_{3,0})^2}$, and third nearest neighbors are at $d_{3,0}=|\mathbf{r}_{1,0}-\mathbf{r}_{3,0}|=
\sqrt{(\frac{a_{1,0}}{2}+\delta_{x,0})^2+(\frac{a_{2,0}}{2})^2+(z_{1,0}-z_{3,0})^2}$.

Ferroelectricity arises from a reduction of structural energy occurring when $d_{3,0}> d_{2,0}$ ($\delta_{x,0}>0$). Distances $d_{2,0}$ and $d_{3,0}$ differ on the relative sign in the first term of the radicand, and can be of similar magnitude (or even equal for Pb-based monochalcogenide monolayers). As complementary order parameters, angles $\alpha_{i,0}$ ($i=1,2,3$) in Figure~\ref{fig:fig1}(a) add up to slightly over 360$^{\circ}$ to the extent that the five atoms defining $\alpha_{1,0}$, $\alpha_{2,0}$ and $\alpha_{3,0}$ nearly but not exactly lie on the same plane as a consequence of the discrete Gauss theorem \cite{prbus,pnas}.

The rhombic distortion angle $\Delta \alpha$ is another geometrical order parameter related to lattice parameters as follows \cite{other4}: $\frac{a_1}{a_2}=\frac{1+\sin \Delta \alpha}{\cos \Delta \alpha}$. $\Delta\alpha$ is greater than zero on the ground state unit cell [Figure~\ref{fig:fig1}(a)].

At finite temperature, atoms forming bonds $d_2$ and $d_3$ will turn ``physical'' or ``temporary'' \cite{jones2002soft}, leading these atomically thin membranes onto a two-dimensional transformation. Two paraelectric ($\delta_x=\delta_y=0$) geometries relevant for these 2D structural transformations are discussed next.

The unit cell in Figure~\ref{fig:fig1}(b) is a square with lattice parameter $a_s$ \cite{Mehboudi2016,other2,other4}. The two added constraints $\delta_x=0$ (for paraelectric behavior) and $a_1=a_2=a_s$ (for a square lattice) reduce the number of independent parameters from five to \textit{three}. The energy of the optimized square unit cell is $E_s=min\{E(a,z_1,z_3)\}=E(a_s,z_{1,s},z_{3,s})$, and its geometry is specified by:
\begin{align}
\label{eq:eq2}
{\bf a}_{1,s}=(a_{s},0,0),\quad {\bf a}_{2,s}=(0,a_{s},0),\qquad\\
\begin{split}
{\mathbf r}_{1,s}=(\tfrac{a_{s}}{2},\tfrac{a_{s}}{2},z_{1,s}),\quad {\bf r}_{2,s}=(0,0,0) &\quad\text{(M)},\\
{\bf r}_{3,s}=(0,0,z_{3,s}),\quad {\bf r}_{4,s}=(\tfrac{a_{s}}{2},\tfrac{a_{s}}{2},z_{1,s}-z_{3,s}) &\quad\text{(X)}. \nonumber
\end{split}
\end{align}
Here $d_{1,s}=|\mathbf{r}_{1,s}-\mathbf{r}_{4,s}|=|\mathbf{r}_{2,s}-\mathbf{r}_{3,s}|=\sqrt{z_{3,s}^2}$, $d_{2,s}=d_{3,s}=|\mathbf{r}_{2,s}-\mathbf{r}_{4,s}|=
\sqrt{(\frac{a_{1,s}}{2})^2+(\frac{a_{2,s}}{2})^2+(z_{1,s}-z_{3,s})^2}$ and all angles (not lying in the same plane) turn into $\alpha_1\gtrsim 90^{\circ}$.  $\Delta\alpha$ is zero on this paraelectric structure \cite{Kai} [Figure~\ref{fig:fig1}(b)]. The energy barrier $J_s=E_s-E_0$ is the energy difference between this square unit cell and the rectangular ground state unit cell; it is the extra energy required to turn from a ferroelectric structure in which a given atom has two second nearest neighbors onto a paraelectric fourfold-symmetric structure where all atoms possess a higher coordination with four second nearest neighbors.

The alternative paraelectric configuration ($\delta_x=0$) shown in Figure~\ref{fig:fig1}(c) has a rectangular unit cell sharing lattice parameters from the ground state structure $a_{1,0}$ and $a_{2,0}$ \cite{fei_prl_2016} thus depending on the \textit{two} free variables $z_1$ and $z_3$ only. It has an energy $E_r=min\{E(a_{1,0},a_{2,0},z_1,z_3)\}=E(a_{1,0},a_{2,0},z_{1,r},z_{3,r})$, leading to a different energy barrier $J_r=E_r-E_0$, and to the following atomistic structure:
\begin{align}
\label{eq:eq3}
{\bf a}_{1,r}=(a_{1,0},0,0),\quad {\bf a}_{2,r}=(0,a_{2,0},0),\qquad\\
\begin{split}
{\mathbf r}_{1,r}=(\tfrac{a_{1,0}}{2}+\delta_{x,0},\tfrac{a_{2,0}}{2},z_{1,r}),\quad {\bf r}_{2,r}=(\delta_{x,0},0,0) &\quad\text{(M)},\\
{\bf r}_{3,r}=(0,0,z_{3,r}),\quad {\bf r}_{4,r}=(\tfrac{a_{1,0}}{2},\tfrac{a_{2,0}}{2},z_{1,r}-z_{3,r}) &\quad\text{(X)}. \nonumber
\end{split}
\end{align}
Neighbor distances now become $d_{1,r}=\sqrt{z_{3,r}^2}$ and $d_{2,r}=d_{3,r}=
\sqrt{(\frac{a_{1,0}}{2})^2+(\frac{a_{2,0}}{2})^2+(z_{1,r}-z_{3,r})^2}$, and there are two angles, $\alpha_1<  90^{\circ}$ and $\alpha_2>  90^{\circ}$.
As it maintains a rectangular unit cell, $\Delta \alpha$ is always nonzero on this paraelectric geometry [Figure~\ref{fig:fig1}(c)].

\subsection{Necessity of corrections for the electronic density of these materials}

Group-IV monochalcogenide monolayers have bonds of varying strengths. The peculiarity of these physical-like bonds becomes apparent by documenting the drastic change of distances (geometry) as these materials are thinned down to monolayers, a situation unheard of in more familiar atomically-thin 2D materials such as graphene and transition metal dichalcogenides, but already realized in black phosphorus \cite{shulen}.

\begin{figure}[tb]
\begin{center}
\includegraphics[width=0.48\textwidth]{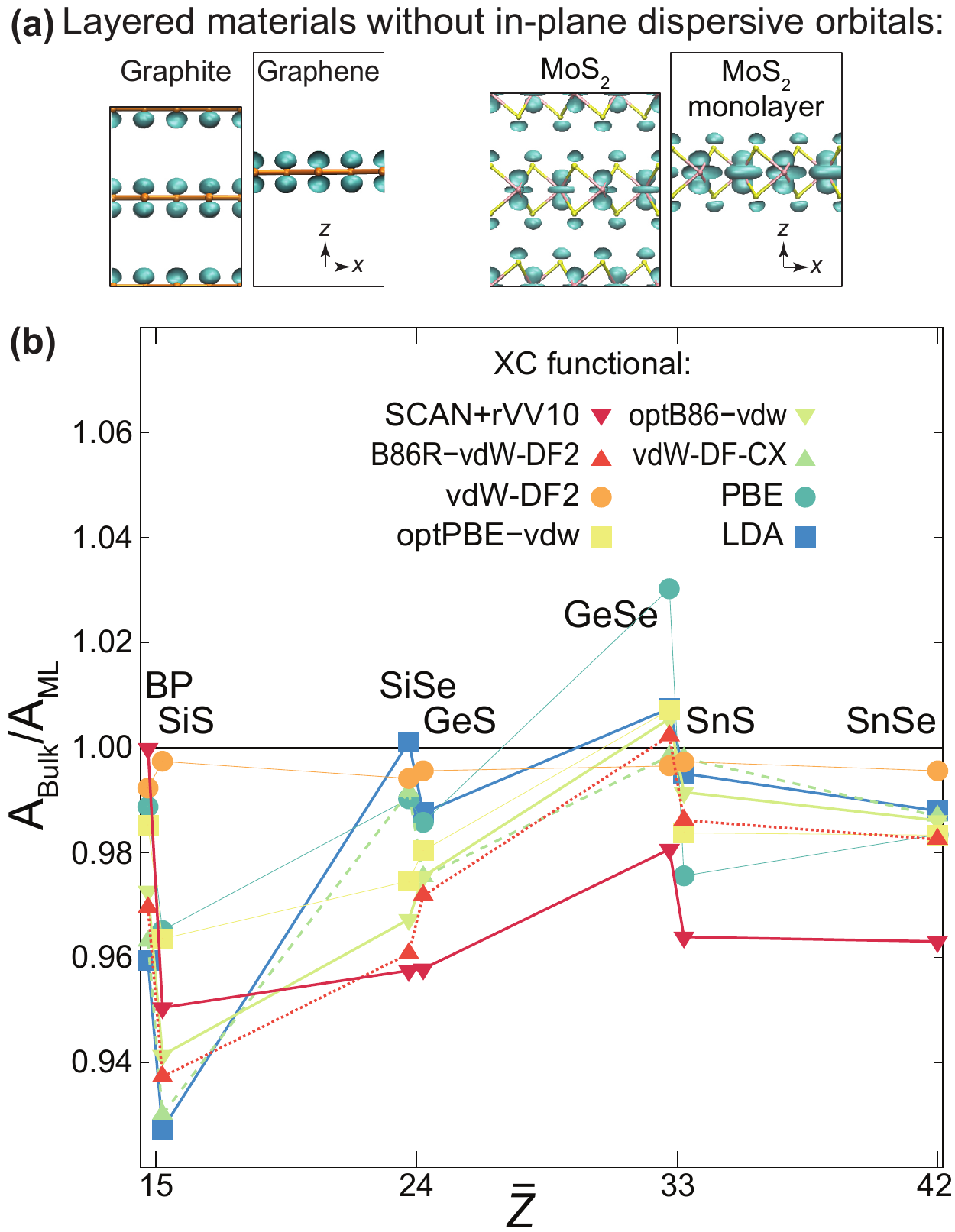}
\caption{(a) Regular layered materials like graphite or bulk MoS$_2$ do not change their area significantly when tinned down to monolayers. (b) In contrast, black phosphorus and most layered group-IV monochalcogenides swell when in monolayer form (i.e., $A_{Bulk}/A_{ML}<0$) due to a lone pair with in-plane components arranging differently in the bulk and in vacuum. (SnTe, GeTe, PbS, PbSe and PbTe, not included, are not layered in the bulk).}\label{fig:fig2}
\end{center}
\end{figure}

Figure~\ref{fig:fig2}(a) displays graphite, graphene, a layered transition metal dichalcogenide (MoS$_2$), and its monolayer. Strong bonds hold individual monolayers together and in direct consequence, the cell area perpendicular to the $z-$direction changes by a negligible amount in going from the bulk onto a monolayer: its ratio, $A_{Bulk}/A_{ML}$, takes values 0.998 to 1.000  for graphite, and 0.996 to 1.003 for MoS$_2$ across the eight XC functionals employed. Interatomic distances are similarly (un)modified.

\begin{figure*}[tb]
\begin{center}
\includegraphics[width=0.96\textwidth]{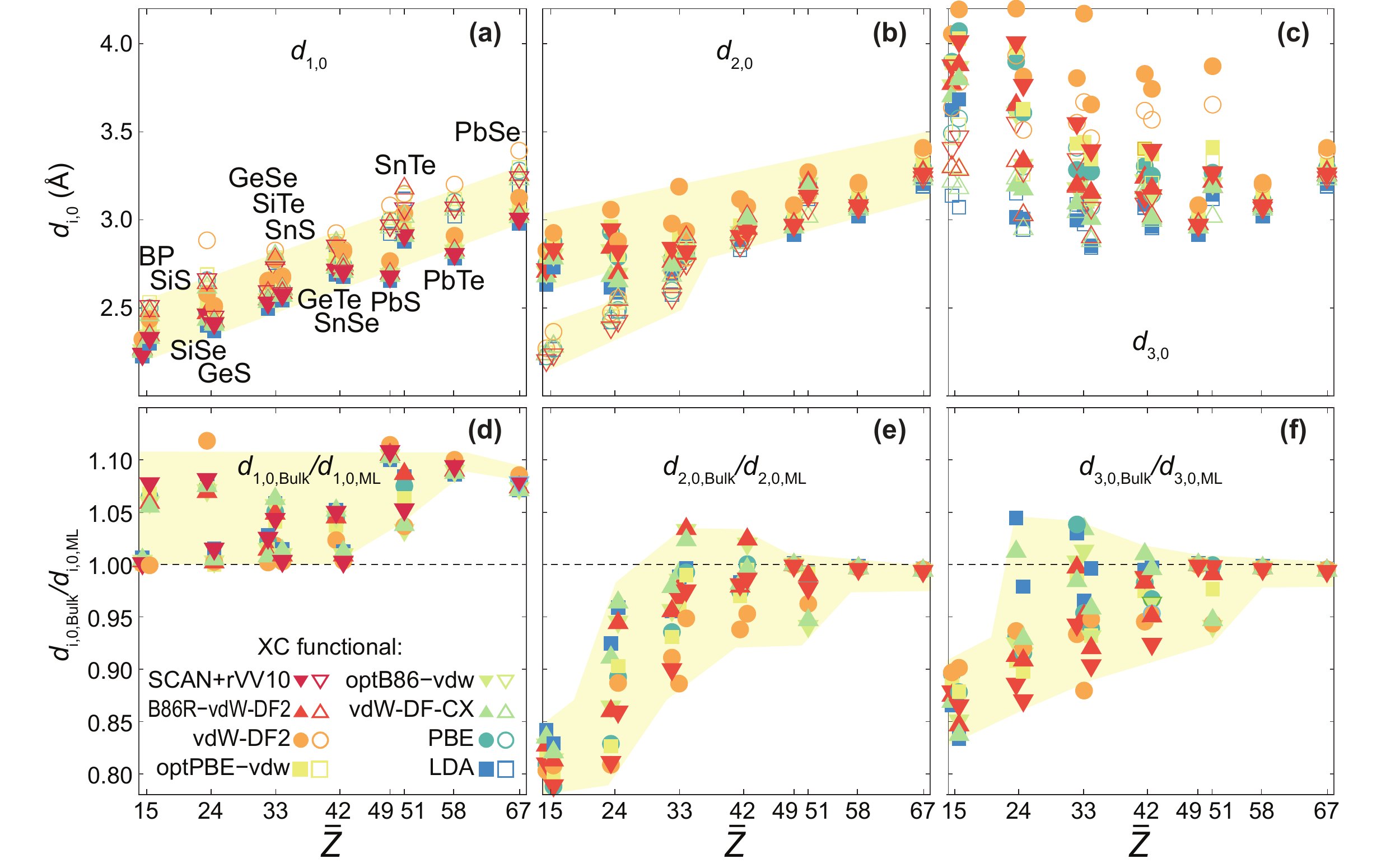}
\caption{(a) $d_{1,0}$; (b) $d_{2,0}$; and (c) $d_{3,0}$ for bulk (open) and monolayers (solid symbols) of black phosphorus and twelve group-IV monochalcogenides. (d), (e) and (f) are their bulk to monolayer ratios. For all XC functionals, $d_1$ is up to 10\% larger in the bulk, while $d_2$ and $d_3$ swell in monolayers by at most as 20\%, a sign of bond plasticity unseen in other 2D materials.}\label{fig:fig3new}
\end{center}
\end{figure*}

\begin{figure*}[tb]
\begin{center}
\includegraphics[width=0.96\textwidth]{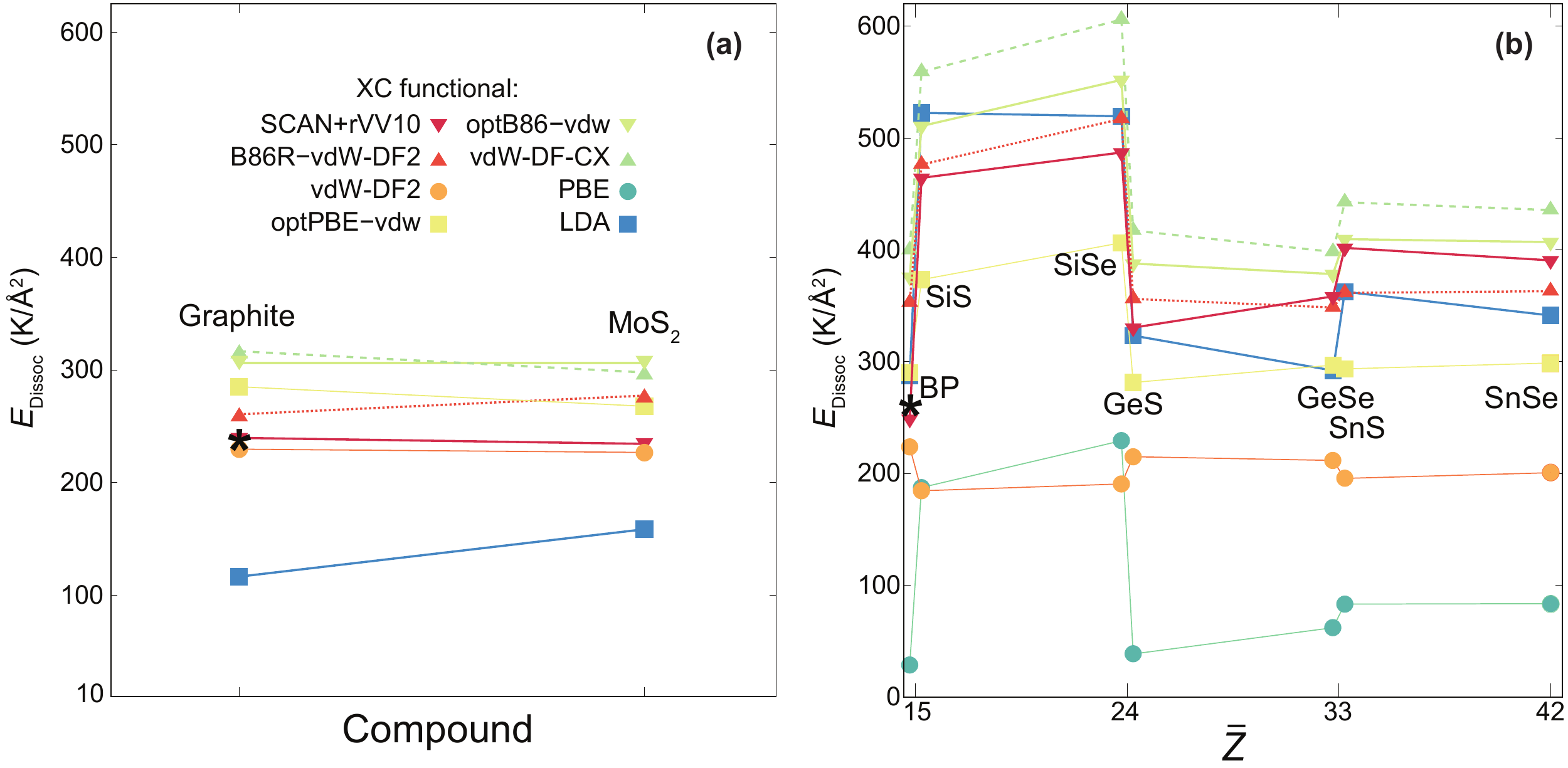}
\caption{(a) $E_{Dissoc}$ for graphite and MoS$_2$. The asterisk corresponds to the value reported in Ref.~\cite{exfoliateGr}. (PBE values turned out to be smaller than 10 K/\AA$^{2}$ and are not shown for that reason.) (b) $E_{Dissoc}$ for group-IV monochalcogenides that are orthorhombic in the bulk. Ge-based
monochalcogenides have a $E_{dissoc}$ comparable to that of black phosphorus, regardless of XC functional. The asterisk corresponds to the QMC data in Ref.~\cite{shulen}.}\label{fig:fig4}
\end{center}
\end{figure*}

In contrast, $A_{Bulk}/A_{ML}$ in Figure~\ref{fig:fig2}(b) has a one order of magnitude larger range--in between 0.930 and 1.040--for these ferroelectric 2D materials. The horizontal axis on Figure~\ref{fig:fig2}(b) is the average atomic number \cite{Mehboudi2016} \footnote{$\bar{Z}=15$ for BP and SiS, $\bar{Z}=24$ for SiSe and GeS, and so on. Datapoints in Figs.~\ref{fig:fig2} to \ref{fig:fig5} were therefore displaced by 0.5 along the horizontal axis just to better differentiate compounds with identical $\bar{Z}$.} $\bar{Z}=(Z_M+Z_X)/2$.  SnTe, GeTe, PbS, PbSe, and PbTe are not layered in the bulk and were not included in this figure for that reason.

We employed a large set of exchange-correlation functionals to explore what distribution of structural parameters is achievable subject to our stringent convergence criteria and numerical results do permit categorizing these approximations. With the exception of GeSe, results obtained with the B86R-vdW-DF2 (in upper red triangles) and optPBE-vdW (in yellow squares) XC
functionals show the closest resemblance to those obtained using the SCAN+rVV10 XC functional, (red inverted triangles). Structures obtained with vdW-DF2 XC functionals (orange circles) show almost no change in area in going from bulk to monolayer. LDA (blue squares) and PBE (cyan circles) XC functionals give ratios that are generally away from the predictions from B86R-vdW-DF2, optPBE-vdW and SCAN+rVV10 XC functionals. All numerical structural and energy data has been made available as Supporting Information.

The fact that $A_{bulk}/A_{ML}$ is mostly smaller or equal to unity in Figure~\ref{fig:fig2}(b), and that interatomic distances change drastically in going from a bulk structure onto a monolayer in Figure~\ref{fig:fig3new}, imply that lone pairs setting layers apart and placing second- and third-nearest neighbors at distances $d_{2,0}$ and $d_{3,0}$ rearrange in going from the bulk to a monolayer (Figure~\ref{fig:fig3new}): \textit{non-covalent bonds} in group-IV monochalcogenides do not just keep layers apart (which is their function in traditional layered materials), they  also \textit{influence the structure within each monolayer}, and require that methods including dispersive corrections for the electronic structure are employed to obtain these monolayers' geometries.

Indeed, while traditional XC functionals like LDA and PBE are good to describe strong (i.e., covalent and ionic) bonding, they are known to fail in describing bonds of intermediate strength accurately. Figs.~\ref{fig:fig2} and \ref{fig:fig3new} indicate 2D structures formed by relatively weak bonds, which unmistakably calls for inclusion of
dispersive corrections.

\subsection{Monolayer dissociation energies}

Exfoliation of these ferroelectric monolayers is work in progress and growth has been a successful route to create and characterize SnTe monolayers on graphitic substrates \cite{Kai,KaiPRL}. A question that has not been addressed yet is whether exfoliating these materials is much more costly than exfoliating graphite, bulk MoS$_2$, or black phosphorus.

Determining an exfoliation energy necessitates creating a thick slab having $N \ge 10$ monolayers, removing one exposed layer, and registering the energy difference to the ($N-1$) stack and an isolated monolayer, while accounting for the thickness-dependent change in area reported in Figure~\ref{fig:fig2}(b). This is a costly process that is compounded by our testing of multiple XC functionals.

With these limitations in mind, we report in Figure~\ref{fig:fig4} the energy difference of a bulk layered compound and its monolayer, i.e., the monolayer dissociation energy (or bulk cleavage energy \cite{shulen})
\begin{equation}\label{eq:diss}
E_{Dissoc}=E_{ML}-\frac{1}{2}E_{Bulk},
\end{equation}
a difference obtained employing the same XC functional in the bulk and monolayer calculations for consistency. The qualitative behavior of $E_{Dissoc}$ in Figure \ref{fig:fig4}(b) for all XC functionals with the exception of vdW-DF2, is as follows: the lowest dissociation energy occurs for
black phosphorus, the Si-based monochalcogenide monolayers have a large dissociation energy, the Ge-based ones have an energy comparable to that of black phosphorus, while Sn-based monolayers have a small increase in dissociation energy relative to those of GeS and GeSe. Due to a well-known underestimation of binding, $E_{Dissoc}$ is the smallest when using the PBE XC functional. The low and almost constant binding across compounds within the vdW-DF2 XC functional (along with its overestimation of structural parameters
documented in the Supplementary Information) leads us to believe it is not suitable to describe these 2D materials. Results obtained with SCAN+rVV10 and optPBE-vdW XC correlation compare favorably.

Across all XC functionals, \textit{Ge-based monochalcogenides have $E_{Dissoc}$ comparable to that of black phosphorus}, and even comparable to the $\sim$300 K/\AA$^2$ obtained for graphene and MoS$_2$ in Figure~\ref{fig:fig4}(a). This is intriguing, considering the lack of successful exfoliation of GeS and GeSe monolayers and that black phosphorous, graphene, and MoS$_2$  exfoliate successfully. The reason for the lack of reports of isolated GeS and GeSe monolayers may be that the mechanical stress produced during exfoliation creates intense local electric fields that violently attract polar molecules and degrade them \cite{asccs}. We hope our results trigger renewed experimental efforts towards the exfoliation of these 2D ferroelectrics and/or new calculations that consider electron-electron correlations more accurately \cite{shulen}.

\subsection{Lattice parameter $a_s$ and thermally accessible elastic energy barriers}

\begin{figure*}[tb]
\begin{center}
\includegraphics[width=0.96\textwidth]{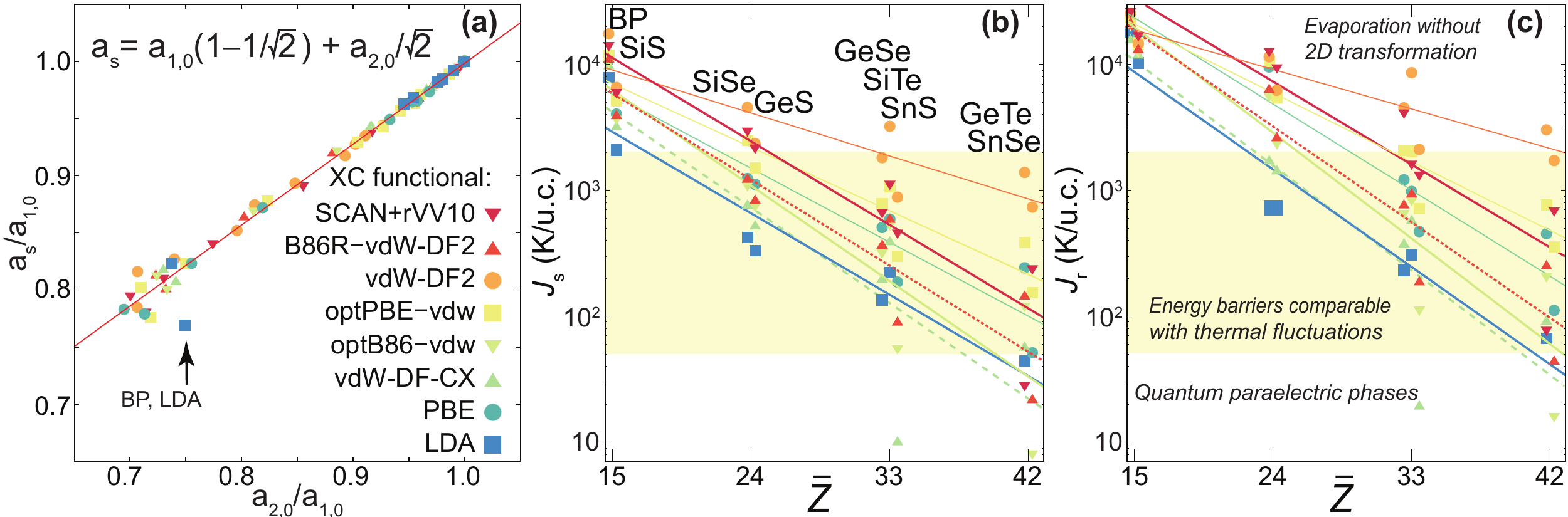}
\caption{(a) Linear relation between $a_s/a_{1,0}$ and $a_{2,0}/a_{1,0}$ for monolayers across XC functionals. (b) $J_s$ decays exponentially with average atomic number and (c) $J_r$ is larger than $J_s$ across these 2D compounds for a given XC functional. Yellow shading in subplots (b) and (c) separates three energy regions: in the lowest one, quantum
fluctuations may forbid the creation of a ferroelectric configuration; the middle one (yellow, from roughly fifty
to a couple thousand Kelvin) is thermally-accessible thus permitting 2D structural transformations; the upper region has barriers too large for 2D structural transformations.}
\label{fig:fig5}
\end{center}
\end{figure*}

Energy barriers may be used to obtain an order of magnitude estimate of the critical temperature $T_c$ at which a ferroelectric to paraelectric two-dimensional transformation takes place \cite{Mehboudi2016,other2,Kai,other4}, thus calling for the reassessment of $J_s$ and $J_r$. To
simplify the determination of $E_s$, one can rely on the linear dependency--regardless of XC approximation--among $a_s$, $a_{1,0}$ and $a_{2,0}$ shown in Figure~\ref{fig:fig5}(a):
\begin{equation}\label{eq:as}
a_s=\left(1-\frac{1}{\sqrt{2}}\right)a_{1,0}+\frac{1}{\sqrt{2}}a_{2,0},
\end{equation}
in which a single outlier corresponds to the black phosphorus monolayer obtained with the LDA XC functional. Figures~\ref{fig:fig5}(b) and
\ref{fig:fig5}(c) show that $J_s$ and $J_r$ decay exponentially with $\bar{Z}$ for all XC functionals. Complete structural parameters necessary to create the  paraelectric structures [Eqs.~(\ref{eq:eq2}) and (\ref{eq:eq3})] are provided as Supporting Materials.

Now, since the square paraelectric unit cell is set by the three variables $a_s$, $z_{1,s}$ and $z_{3,s}$ and the rectangular paraelectric structure by the two parameters $z_{1,r}$ and $z_{3,r}$, we assert that the increased number of structural constraints on the rectangular paraelectric structure underpins its higher elastic energy barrier. This is indeed the case, and our compiled numerical data indicates that
\begin{equation}\label{eq:jsjr}
1.4 \le \frac{J_r}{J_s}\le 7.7, \nonumber
\end{equation}
making the \textit{square unit cell} in Figure~\ref{fig:fig1}(b) the lowest energy, \textit{preferred paraelectric geometry}. We note here that $J_s=J_r=0$ for these (mostly Pb-based) compounds whose ground state unit cell is square already.

Showcasing the SnSe monolayer again, $J_s$ ($J_r$) is as follows: 738.4	(1722.2), 243.5	(702.6), 154.8	(356.2), 51.3 (111.5), 21.3 (43.0), 8.3	(16.4) in units of K/u.c. for
vdW-DF2, SCAN+rVV10, optPBE-vdW, PBE, B86R-vdW-DF2, and optB86-vdW XC functionals, respectively, and zero when using the LDA and vdW-DF-CX XC functionals.
Note, once again, the closest correspondence for $J_s$ and $J_r$ occurring when the SCAN+rVV10 and optPBE-vdW XC functionals are employed, that PBE XC underestimates these barriers when compared to these two functionals, and that these barriers take on their maximum values when using the DF2-vdW XC functional.

Crucially, some elastic energy barriers are of the order of thermal fluctuations, making these materials atomically thin soft membranes. To emphasize this point, the soft
yellow shading in Figs.~\ref{fig:fig5}(b) and \ref{fig:fig5}(c) separates three energy regions: at low values of $J_s$ or $J_r$, the energy required to switch among ferroelectric ground states could become accessible via quantum fluctuations that would then prevent the creation of a ferroelectric structure, in a situation similar to a two dimensional quantum paraelastic phase \cite{arxiv,prb}. Within the yellow shading, the barrier is thermally accessible and 2D phase transformations become possible; GeSe, SnSe \cite{Mehboudi2016,other2,other4} and SnTe \cite{Kai,sntebl} monolayers belong to this category. Barriers are so high in the upper unshaded region that these 2D materials are stiff enough not to be significantly perturbed by thermal or quantum fluctuations and can only evaporate. The black phosphorus monolayer belongs to this latter category.

Interactions with a supporting substrate (e.g., Refs.~\cite{Kai,KaiPRL}) will necessarily modify these elastic energy barriers. Nevertheless, the structural complexity even in the freestanding configuration being uncovered here requires reaching agreements as to how to understand these freestanding structures, before complications introduced by supporting substrates are considered. Similar distinctions on the phenomenology of supported \textit{versus} unsupported 2D materials have been considered before \cite{Halperin}.

The spread of data when comparing the eight studied XC functionals and the lack of experimental exfoliation and characterization of freestanding monolayers prevents selecting an optimal functional. Discarding vdW-DF2 due to the reasons presented above, the optPBE-vdW XC functional gives results similar to those obtained using the SCAN+rVV10 XC functional for structure, energy barriers [Figure~\ref{fig:fig5}(b,c)] and dissociation energy (Figure~\ref{fig:fig4}). Verification of the best functional for these ferroelectric monolayers will require experimental data on freestanding samples or more precise calculations that consider electron-electron interactions explicitly.

\section{Conclusions}\label{sec:conclusion}
The following conclusions are reached as a result of this work:
\begin{enumerate}
\item{}Similar to black phosphorus, the in-plane unit cell of these 2D ferroelectrics changes its area in going from the bulk to a monolayer.
\item{}We realize a hierarchy of independent quantities that uniquely define a ground ferroelectric unit cell and square and rectangular paraelectric unit cells; this hiearchy works for group-IV monochalcogenides. (For black phosphorus, pairs of heights are equal, so these monolayers are described by four, two, and one independent parameters on the ferroelastic ground state, the square paraelastic, and the rectangular paraelastic unit cells, respectively.)
\item{}We provide the dissociation energy of these materials is about three times larger than the energy required to exfoliate graphite and MoS$_2$.
\item{}We show a simple, linear, relation among the square paraelectric unit cell and the lattice parameters of the rectangular ferroelectric ground state unit cell.
\item{}We establish that the reduced number of independent structural variables correlates with larger elastic energy barriers on a rectangular paraelectric unit cell when compared to the barrier of a square structure. This has consequences on the expected transition temperatures on these two different (NPT versus NVT) scenarios.
\end{enumerate}

\section{Acknowledgments}
We acknowledge conversations with Kai Chang, Benjamin Fregoso, and Haowei Peng, and thank B. J. Miller for technical assistance. S.P.P. was funded by the U.S. National Science Foundation (Grant No. DMR-1610126); J.W.V and S.B.L. by an Early Career Grant from the U.S. Department of Energy, Office of Basic Energy Sciences (Award  DE-SC0016139). Calculations were performed on Cori at NERSC, a U.S. DOE Office of Science User Facility under Contract No. DE-AC02-05CH11231, and at the University of Arkansas' Trestles and Pinnacle supercomputers, funded by the U.S. National Science Foundation (Grants 0722625, 0959124, 0963249, and 0918970), a grant from the Arkansas Economic Development Commission, and the Office of the Vice Provost for Research and Innovation.


%

\end{document}